\documentstyle[12pt]{article}
\def\be{\begin{equation}}
\def\ee{\end{equation}}
\def\bq{\begin{eqnarray}}
\def\eq{\end{eqnarray}}
\def\ra{\rightarrow}
\def\n{\nonumber}
\def\g{\gamma}
\def\vp{\varphi}

\begin{document}
\input{psfig}

\begin{flushright}
SPhT-t97/ \\
July 1997
\end{flushright}
\vspace{1cm}
\begin{center}
{\bf Pion Light-Cone Wave Functions  and Light-Front Quark Model}
\end{center}
\begin{center}{V.M. Belyaev$^*$ }
\end{center}
 \begin{center}
 {\em SPhT, CEA-SACLAY, 91191 Gif-sur-Yvette, CEDEX, France}
\end{center}
\begin{center}{Mikkel B. Johnson}  \\
{Los Alamos National Laboratory, Los Alamos, NM 87545, USA
}
\end{center}
\vspace{1cm}
\begin{abstract}
We discuss a relation between the light-front quark model and QCD. We argue 
that this model can be used for an evaluation of the light-cone wave functions 
for moderate values of $u$, but that it is inapplicable
for this purpose in the region near the end points $u=(0,1)$.  
We find additional support for a recent analysis in which it was 
claimed that the twist-two pion wave function attains its asymptotic form.  The 
asymptotic twist-four two-particle wave function is also in good agreement 
with the light-front quark model.

\vspace{0.5cm}

\noindent PACS number(s): 11.15.Tk, 12.38.Lg, 12.39.Ki, 12.39.-x, 14.40.Aq
\end{abstract}

\vspace{1cm}
\flushbottom{$^*\overline{On\;  leave\;  of\;  absence }
\;from\;  ITEP,\;  117259\;  Moscow,\; Russia.$}

 \newpage

\section{Introduction}

The light-front quark model (LFQM)  \cite{mvt1},  is based on 
the algebra of the generators of the Lorentz-group in light-front 
dynamics \cite{mvt2}.  The specific 
dynamical input is made through a parametrized form for the quark wave 
function.  With specific choices of the parameters, the model describes 
numerous hadron properties, including form factors for $Q^2\sim$ 1 
 GeV$^2$ (see for example \cite{rcqm}).  Thus, this model establishes a 
phenomenological link between hadron 
properties and the wave function of the quark constituents that has been 
successful in many instances.  

It is desirable to investigate a deeper theoretical connection between the 
LFQM wave function and QCD, not only to establish 
more firmly the dynamical basis of the LFQM but also with an eye to 
improve the general understanding and interconnections among various approaches.
In our work, we derive a connection between the LFQM and
QCD light-cone wave functions by equating 
the matrix element
$<0|\bar{d}(0)\g_\mu \g_5u(x_1)|\pi^+(P)>$ in the two appproaches. This leads 
to an expression for the light-cone QCD wave function in terms of the 
LFQM wave function.

Some relevant questions were considered in  \cite{zh}.  Particulary, 
in \cite{zh} it was noted that the asymptotic behavior $u\ra (0,1)$ 
of the QCD light-cone wave function 
can not be reproduced within a constituent quark  model (CQM) 
with equal-time wave functions
\cite{iz} $\psi_{CM}( q^2)\sim \exp(- q^2)$, which can
be represented in the form of the light-cone (LC) wave function 
by indentification (see \cite{bhl}):
\bq
 \vec q^2\leftrightarrow
\frac{ k_\perp^2+m^2}{4u(1-u)}-m^2,\;\;\;\psi_{CQM}(\vec q^2)
\leftrightarrow\psi_{LC}\left(
\frac{ k_\perp^2+m^2}{4u(1-u)}-m^2
\right),
\label{b1}
\eq
where $m\simeq 300MeV$ is the constituent mass.

The same conclusion is reached in the case of the LFQM \cite{sim}.  Therefore, we can not 
expect to obtain a good description of the QCD light-cone wave function in 
terms of the quark model near the end points $u=(0,1)$.  In the next section 
we  discuss the reason for this disagreement. However, in most cases hadronic 
matrix elements calculated in the quark model are saturated at moderate $u$, 
and the contribution of the region near the end points is small.  In this case, 
one can expect that the LFQM can describe the light-cone wave functions for 
moderate $u$.

As mentioned above, the LFQM is able to describe hadron matrix elements even 
for large values of momentum transfer:  $Q^2\sim 1 GeV^2$. This means that the 
LFQM gives a good description of hadron matrix elements at small 
distances $x^2\ll \Lambda_{QCD}^{-2}$.  Therefore, it is reasonable to use such a 
model for the evaluation of matrix elements of operators defined at a scale 
$\mu\sim 1 GeV\gg \Lambda_{QCD}$.  Below, we will assume that the matrix 
elements (which depend on the normalization point)  are defined at the 
scale $\mu\sim 1GeV$.

In this paper we consider the transition amplitude of a pion to the vacuum by 
a nonlocal gauge invariant operator of the axial current. This amplitude can be 
expressed in terms of QCD light-cone wave functions \cite{cz,er,bl}.
Assuming that the LFQM can give a good decription for this type of 
amplitude (except the region near the end points $u=(0,1)$, as discussed 
above) we  express the QCD light-cone wave functions of a pion  as an integral 
over the LFQM wave function.  Using the published parametrizations \cite{wj} 
for the latter, we show  that the twist-2 pion wave function is very close to 
its asymptotic form. This observation confirms the result obtained 
in \cite{bj1} based on constraints for the light-cone wave function determined 
from QCD sum rules\cite{SVZ} for $g_{\pi NN}$ coupling 
constant \cite{bk,bf} and from the QCD sum rule for the pion structure 
function \cite{bj2,bj2a}.  The constraint  obtained in \cite{rr} also 
indicates that the the light-cone wave function is close to its asymptotic form.

In the LFQM, we also evaluate the two-particle twist-4 light-cone pion wave 
function.  We find the asymptotic form for the twist-4 light-cone wave 
function to be in good agreement with the LFQM description.

\section{Pion Wave Function}

In our work, we derive a connection between LFQM and
QCD light-cone wave functions by equating 
the gauge invariant matrix element
\bq
<0|\bar{d}(0)\g_\mu U[0,x]\g_5u(x)|\pi^+(P)>
\label{1}
\eq
 in the two appproaches,
where $U[0,x]={\cal P}e^{ig\int_x^0A_\mu(y)dy_\mu}$.
Note that in LFQM considered here the gauge fields are not
taken into account. Nevertheless we can assume that
in the LFQM the quark wave
fuctions correspond to the fixed-point gauge $x_\mu A_\mu(x)=0$.
Then $U[0,x]=1$ and the matrix element (\ref{1})  can be evaluated in
terms of the constituent quark wave functions.

In  QCD, the twist-2 and twist-4 two-particle wave functions of the
pion are defined by the following matrix element:
\bq
&   &<0|\bar{d}(0)\g_\mu \g_5u(x)|\pi^+(P)> \nonumber \\
& = & if_\pi P_\mu\int_0^1e^{-iu(Px)}[\vp_\pi(u)+x^2g_1(u)+x^2G_2(u)
+O(x^4)]du 
\nonumber \\
& + &f_\pi x_\mu\int_0^1e^{-iu(Px)}g_2(u)du+O(x^2)
\label{vpp}
\eq
where $\vp_\pi(u)$ is the twist-2 wave function, where $g_1(u)$ and 
$g_2(u)$ are the twist-4 pion wave functions, where 
and $G_2(u)=-\int_0^ug_2(u)du$.

It is useful to first examine the situation where the quark field $u(x_1)$ is 
placed at the origin, in which case
\bq
<0|\bar{d}(0)\g_\mu\g_5u(x_1=0)|\pi^+(P)>=iP_\mu f_\pi  ,
\label{vp}
\eq
where 
\bq
\bar{d}(x_2=0)=\int e^{ip_2^\prime\cdot x_2}\mid_{x_2=0}\bar{d}(p_2^\prime )
\frac{d^3p_2^\prime }{(2\pi)^3}
\label{vp1}
\eq
and
\bq
u(x_1=0)=\int e^{-ip_1^\prime \cdot x_1} \mid_{x_1=0}u(p_1^\prime )
\frac{d^3p_1^\prime }{(2\pi)^3}.
\label{vp2}
\eq 
Following analysis of Ref.\cite{wj}, one can determine from Eq.(\ref{vp}) 
the relationship of the pion decay constant to the LFQM wave function.
Substituting Eqs.(\ref{vp1},\ref{vp2}) into Eq.(\ref{vp}), introducing 
the LFQM pion wave function $\psi$ defined in terms of the S-wave orbital wave 
function $\phi (p)$,
\bq
\phi(p)=\sqrt{\frac{(2\pi)^3}{N_c}}\frac{1}{\pi^{3/4}\beta^{3/2}}
e^{-\frac{{p}^2}{2\beta^2}} ,
\label{pwf}
\eq
using the analysis of Ref.\cite{wj}, and changing variables to 
the total and relative momenta, one finds that ($N_c=3$)
\bq
f_\pi=\frac{\sqrt{3}}{\pi^{5/4}}\frac{m}{\beta^{3/2}}\int_0^\infty
\frac{p^2e^{-\frac{{p}^2}{2\beta^2}} dp}{(p^2+m^2)^{3/4}},
\label{vp4}
\eq
where $m$ is the constituent mass of a $u$ or $d$ quark. 
In Ref.\cite{wj} it was suggested to use the following set
of parameters:
$\beta=0.3194GeV$ and $m=0.25 GeV$ \cite{wj};
this gives  $f_\pi=0.130 GeV$.

To draw a correspondence with Eq.(\ref{vpp}), let us consider the case 
$x_1=x$  in Eq.(\ref{vp}).  It is then clear that the $u$-quark wave function
gets a phase factor $e^{ip_1x}$.  In terms of light-cone variables,
\bq
p_1 &=& (\zeta P^+, p_1^-,\vec p_\perp)
\n
\\
P^\pm &=& P^0\pm P^3
\n
\\
p_1^\pm &=& p_1^0\pm p_1^3,
\label{bb1}
\eq
the phase factor has the following form:
\bq
e^{ip_1x}=e^{-i\zeta P^+x^-+i\vec p_\perp\vec x_\perp} ,
\label{bb3}
\eq
where by definition $p_1^+=\zeta P^+$.  In Eq.(\ref{bb3}) we have dropped the 
$p^-$ term, consistent with the fact that in the LFQM the $p^-$ dependence is 
usually not considered \cite{mvt1}.  This is justified as long 
as $\zeta \ne 0$ since, in the infinite momentum frame ($P^+\ra\infty$), the 
dependence on $p_1^-$,
\bq
p_1^-=\frac{\vec p_\perp^2+m^2}{\zeta P^+} ,
\label{bb2}
\eq
is suppressed.  However, the suppression is absent in the limit when $\zeta\ra 
0$.  In the region of small $\zeta$ it is incorrect to neglect the 
$p^-$ dependence, and one therefore cannot apply the LFQM for calculations of 
matrix elements here.  It is clear that one draws the same conclusion for 
matrix elements where the dominant contribution comes from the region 
$\zeta\simeq 1$.  It is nevertheless possible to make a comparison to the 
light-front wave functions at $x^+=0$, being mindful of the concerns in the 
region of the end points, $\zeta\simeq (0,1)$.  These remarks are  
useful for understanding of limits of LFQM applicability.

Then, from Eqs.(\ref{vpp},\ref{bb3}) one can find that
\bq 
<0|\bar{d}(0)\g_+ \g_5u(x)|\pi(P)> \nonumber \\
 =  iP_+\frac{\sqrt{3}}{\pi^{5/4}}\frac{m}{\beta^{3/2}}
\frac{1}{4\pi}\int
\frac{e^{-\frac{{p}^2}{2\beta^2}}d^3p}{(p^2+m^2)^{3/4}}
e^{-ip_1 \cdot x},
\label{vpg}
\eq
where we work in the frame of reference in which ${P}_\perp=0$. 
The integral over $p_1^+$ can be rewriten in the form of the integral over
the parameter $\zeta$, and the Jacobian of the transformation is
\bq
J=\frac{\sqrt{p_\perp^2+m^2}}{[4\zeta (1-\zeta )]^{3/2}} .
\label{jac}
\eq
As a result of this transformation, Eq.(\ref{vpg}) becomes:
\bq
& &  \frac{iP_+}{\pi}\frac{\sqrt{3}}{\pi^{5/4}}\frac{m}{\beta^{3/2}}
\int_0^1
\frac{e^{-i\zeta P^+x^-}d\zeta}{[4\zeta (1-\zeta)]^{3/4}}\int 
\frac{d^2p_\perp e^{-\frac{1}{2\beta^2}[\frac{p_\perp^2+m^2}{4\zeta (1-\zeta)}
-m^2]}}{(p_\perp^2+m^2)^{1/4}}
e^{i{p}_\perp\cdot{x}_{\perp}} \nonumber \\
& = & if_\pi P^+\int_0^1e^{-iu(P^+x^-)}[\vp_\pi(u)-x_\perp^2g_1(u)-
x_\perp^2G_2(u)]du.
\label{lqf}
\eq
Here we use the fact that $x^2=-x_\perp^2$ when $x^+=0$.
>From Eq.(\ref{lqf}) it follows that $\zeta=u$.

Using Eq.(\ref{lqf}) and expanding this expression up to the terms $x_\perp^2$
we obtain the following formulae for the twist-2 and twist-4
light-cone wave functions:
\bq
&  & f_\pi \vp_\pi(u) \nonumber \\
& =  & \frac{\sqrt{3}m}{\pi^{5/4}\beta^{3/2}
{[4u (1-u)]^{3/4}}}\int_0^\infty 
\frac{dp_\perp^2 e^{-\frac{1}{2\beta^2}[\frac{p_\perp^2+m^2}{4u(1-u)}
-m^2]}}{(p_\perp^2+m^2)^{1/4}}
\label{lqff1}
\eq
and 
\bq
&  & f_\pi [g_1(u)+G_2(u)] \nonumber \\
& =  &  \frac{1}{4}\frac{\sqrt{6}m}{\pi^{5/4}\beta^{3/2}
{[4u (1-u)]^{3/4}}}\int 
\frac{dp_\perp^2 e^{-\frac{1}{2\beta^2}[\frac{p_\perp^2+m^2}{4u(1-u)}
-m^2]}}{(p_\perp^2+m^2)^{1/4}}
{p}_\perp^2,
\label{lqff2}
\eq
where the additional factor of $1/4$ in Eq.(\ref{lqff2}) arises from the 
expansion of the exponential to second order and subsequent integration over 
the square of the angle between ${p}_\perp$ and $x_\perp$.  

It was noted above that we can not apply the LFQM to determine matrix elements 
having their dominant contribution occurring near the end points $\zeta=(0,1)$. 
Therefore, since $\zeta=u$, we cannot expect a good description of the QCD 
light-cone wave functions in terms of the LFQM in the region near the end 
points $u=(0,1)$.

\section{Numerical Results}

The integrals in Eqs.(\ref{lqff1},\ref{lqff2}) may be 
 expressed  in terms of the incomplete
Gamma function \cite{as} $\Gamma(a,x)=\int_x^\infty dz e^{-t} t^{a-1}$.  
The result is
\bq
f_\pi \vp_\pi(u)=\frac{2^{1/4}\sqrt{6}}{\pi^{5/4}}me^{\frac{m^2}{2\beta^2}}
\Gamma(\frac{3}{4},z_0)
\label{fr1}
\eq
and
\bq
f_\pi [g_1(u)+G_2(u)]=\frac{2^{1/4}\sqrt{6}}{4\pi^{5/4}}e^{\frac{m^2}{2\beta^2}}
m^3[\frac{1}{z_0}\Gamma(\frac{7}{4},z_0)-\Gamma(\frac{3}{4},z_0)] ,
\label{fr2}
\eq
where $z_0=\frac{m^2}{8\beta^2u(1-u)}$.

We compare our results with asymptotic QCD light-cone wave functions:
\bq
\vp_\pi(u)&=&6u\bar{u} \nonumber \\
g_1(u)&=&\frac{5}{2}\delta^2u^2\bar{u}^2 \nonumber \\
g_2(u)&=&\frac{10}{3}\delta^2\bar{u}u(u-\bar{u}) \nonumber \\
G_2(u)&=&\frac{5}{3}\delta^2\bar{u}^2u^2 ,
\label{asym}
\eq

where $\bar{u}=1-u$, and with constraints on the light-cone wave function
obtained from the QCD light-cone sum rule for the $g_{\pi NN}$ coupling 
constant \cite{bf} and from the light-cone QCD sum rule for the pion structure 
function \cite{bj2}.  
The QCD sum rule estimate \cite{nsvz} for $\delta^2$ yields $\delta^2=0.2$ 
GeV$^2$.  Comparing
Eq.(\ref{asym}) to Eqs.(\ref{fr1},\ref{fr2}), we find the results 
shown in Fig.1 and Fig.2, respectively.

We see that the agreement for the twist-2 wave function is about 10\% and for 
the twist-4 wave function is 12 \% at the peaks, $u=0.5$.  The LFQM predicts
a broader distribution  and is consistent with results of Ref.\cite{bj1}.

\section{Discussion and Conclusion}

We have obtained a  connection between the LFQM wave function of the pion and 
the corresponding two-particle wave functions in QCD by identifying matrix 
elements of quark fields, $<0|\bar{d}(0)\g_\mu \g_5u(x_1)|\pi^+(P)>$.  This 
connection has permitted us to check conclusions obtained from the QCD sum 
rule analysis for the pion against findings in the LFQM.  We also found a 
simple reason to be wary of the comparison between the LFQM predictions 
and the QCD light-cone wave functions in the region where $u=(0,1)$: one of the 
main LFQM results, namely the absence of a sensitivity to the $p^-$ component 
of the quark momentum, does not permitt us to take seriously the comparisons 
near the end points $u=(0,1)$.  It was shown that the LFQM description of QCD 
light-cone wave functions indicates not only that the twist-2 pion light-cone 
wave function is close to its asymptotic form but also that the form of the 
twist-4 light-cone wave function $g_1(u)+G_2(u)$ is not far from asymptotic.

\section{Acknowledgements}
One of the authors (V.B.) thanks I.Narodetsky for stimulating
discussions.
This research was sponsored in part by the U.S. Department of Energy
at Los Alamos National Laboratory under contract 
W-7405-ENG-36 and by the INTAS  Grant 93-0283 and by the
CRDF Grant RP-2-132.


\begin{thebibliography}{99}

\bibitem{mvt1} M.V. Terent'ev, Yad. Fiz. {\bf 24} (1976) 207 [Sov. J. Nucl.
Phys. {\bf 24} (1976) 106]; V.B. Berestetsky and M.V. Terent'ev, {\it ibid} 
{\bf 24} (1976) 1044; [{\bf 24} (1976) 547] {\bf 25} (1977) 653 [{\bf 25} 
(1977) 347].
\bibitem{mvt2} M.V. Terent'ev, Yad. Fiz. {\bf 38} (1983) 213 [{\bf 38} 
(1983) 124].
\bibitem{rcqm} I.G. Aznaurian, A.S. Bagdasarian, and N.L. Ter-Isaakian,
Phys. Lett. {\bf B112} 393; I.G. Aznauryan and K.A. Oganessyan, Phys. Lett.
{\bf B249} (1990) 309; P.L. Chung, F. Coester, and W.N. Polyzou,
Phys.Lett. {\bf B205} (1988) 545;
 P.L. Chung, F. Coester, B.D. Keister, and W.N. Polyzou,
Phys.Rev. C {\bf 37} (1988) 2000;
F. Cardarelli, I.L. Grach, I.M. Narodetskii, G. Salme, and S. Simula,
Phys.Lett {\bf B359} (1995) 1;
F. Cardarelli, I.L. Grach, I.M. Narodetskii, E. Pace, G. Salme, and S. Simula,
Phys.Rev. {\bf D53} (1996) 6682.
\bibitem{zh} A.R. Zhitnitsky, in Minneapolis 1996,
Continuous advances in QCD, 345-357; E-print Archive:
hep-ph/9605226.
\bibitem{iz} N. Isgur, in The new Aspects of Subnuclear Physics,
edited by A. Zichichi (Plenum Publishing Corporation, New York, 1980);\\
J. Rosner, in Techniques and Concepts of High Energy
Physics, edited by T. Ferbel (Plenum Publishing Corporation, New York, 1981).
\bibitem{bhl} S.J. Brodsky, T. Huang and P. Lepage,
in Particles and Fields, edited by A.Z. Capri and A.N. Kamal (Plenum
Publishing Corporation, New York, 1983).
\bibitem{sim} V.L. Morgunov, V.I. Shevchenko, and Yu.A. Simonov,
E-Print Archive: hep-ph/9704282.
\bibitem{cz} V.L. Chernyak and A.R. Zhitnitsky, Phys.Rep. {\bf 112} (1984) 173.
\bibitem{er}A.V. Efremov and A.V. Radyushkin, Phys.Lett. {\bf B94} (1980) 245.
\bibitem{bl} G.P. Lepage and S.J. Brodsky, Phys. Rev. D {\bf 22} (1980) 2157.
\bibitem{wj} W. Jaus, Phys. Rev. D {\bf 44} (1991) 2851.
\bibitem{bj1} V.M. Belyaev and M.B. Johnson, LA-UR-97-1119, SPhT t97/032 and 
E-Print Archive: hep-ph/9703244.
\bibitem{SVZ} M.A. Shifman, A.I. Vainshtein, and V.I. Zakharov, Nucl. Phys.
{\bf B147} (1979) 385, 448.
\bibitem{bk}I.I. Balitsky, V.M. Braun, and A.V. Kolesnichenko, Yad.Fiz. 
{\bf 41} (1985) 282.
\bibitem{bf}V.M. Braun and I.E. Filyanov, Z.Phys. {\bf C44} (1989) 157.
\bibitem{bj2} V.M. Belyaev and M.B. Johnson, To be published in Phys. Rev.
D; LA-UR-97-1118, SPhT t97/010 and E-Print Archive: hep-ph/9702207.
\bibitem{bj2a}V.M. Belyaev and M.B. Johnson, 
 in Minneapolis 1996,
Continuous advances in QCD, 338-344; E-print Archive:
hep-ph/9605279.
\bibitem{rr} A.V. Radyushkin and R. Ruskov, Nucl.Phys. {\bf B481} (1996) 625.
\bibitem{as} M. Abramowitz and I.A. Stegun, {\it Handbook of Mathematical
Functions} (Dover Publications, Inc., New York, 1965).
\bibitem{nsvz} V.A. Novikov, M.A. Shifman, A.L. Vainshtein, and V.I
Zakharov, Nucl. Phys. {\bf B237} (1984) 525.
\newpage

{\large \bf Figure Captions}\\ \\
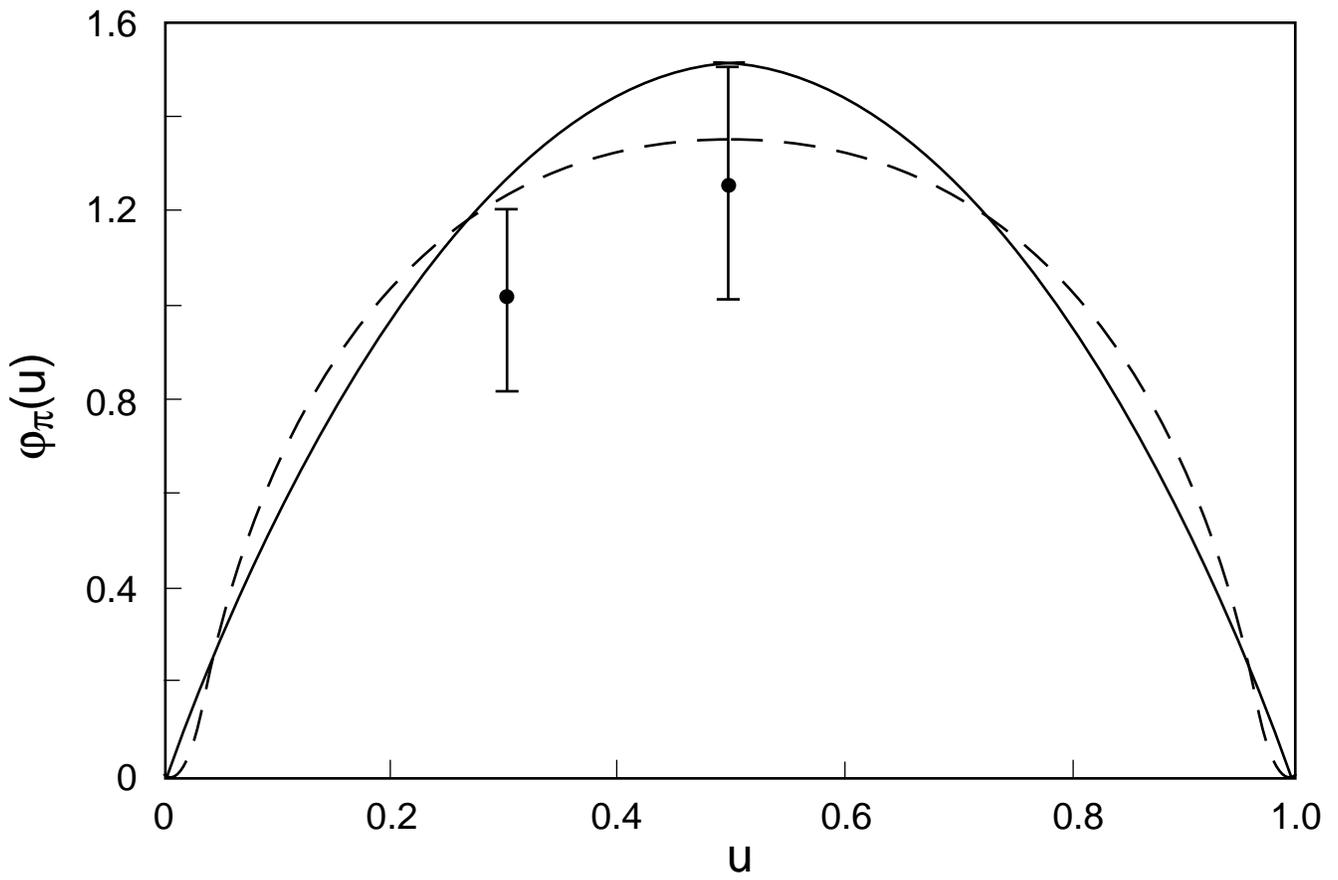
Figure 1.
Comparison between the asymptotic twist-2 light-cone wave function
$\vp_\pi(u)$ (solid curve) and the result of the LFQM (dashed curve), given 
by Eq.(\ref{fr1}). The points at $u=0.3$ and $u=0.5$ are 
constraints determined in Refs.\cite{bj2} and \cite{bf}, respectively.\\ \\
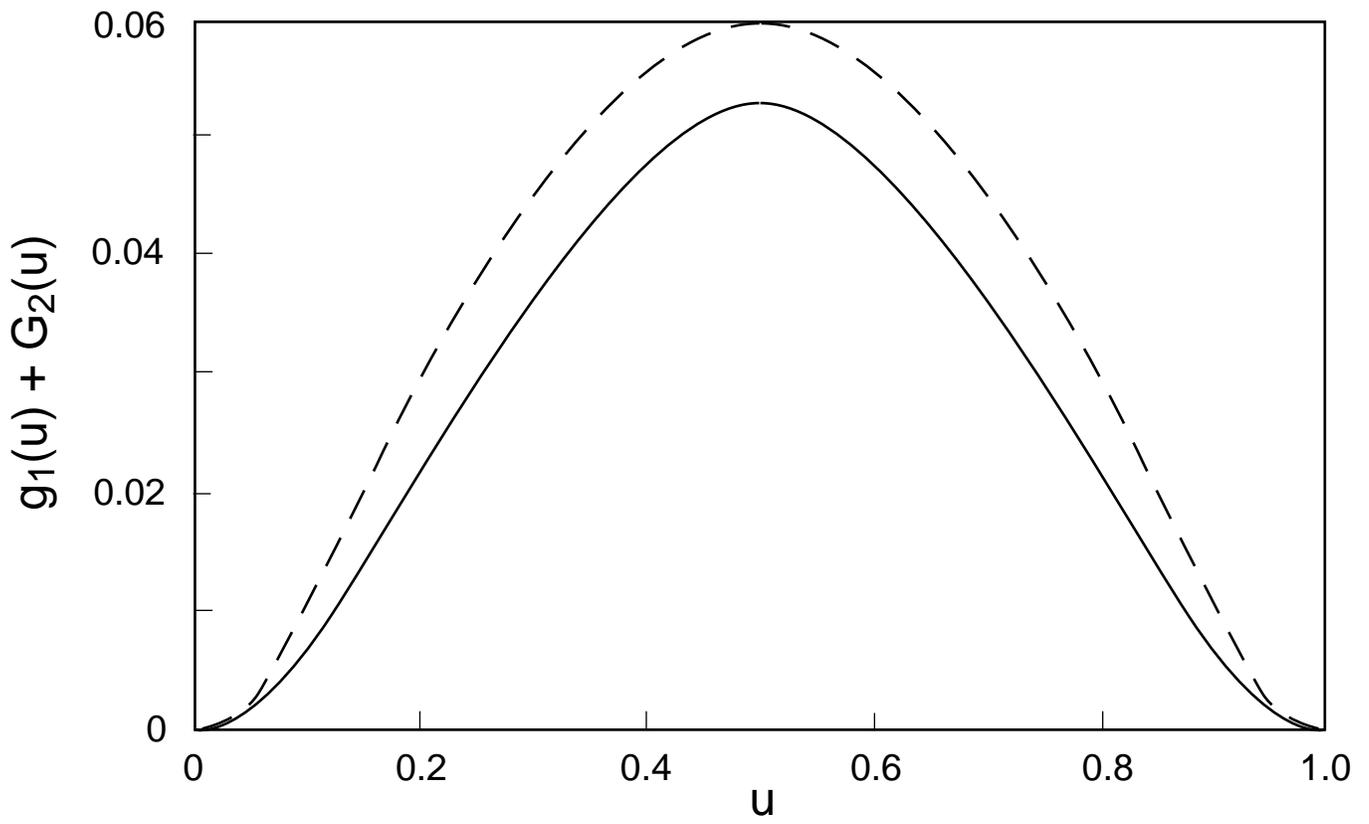
 Figure 2. Comparison between the asymptotic twist-4 light-cone wave function
$g_1(u)+G_2(u)$ (solid curve) and the result of the LFQM (dashed curve), 
given by Eq.(\ref{fr2}).


\end{thebibliography}
\end{document}